\documentclass{article}

\PassOptionsToPackage{numbers,sort&compress}{natbib}

\usepackage[final]{hrl_nips_2017}

\usepackage[utf8]{inputenc} 
\usepackage[T1]{fontenc}    
\usepackage{hyperref}       
\usepackage{url}            
\usepackage{booktabs}       
\usepackage{amsfonts}       
\usepackage{nicefrac}       
\usepackage{microtype}      

\usepackage{graphicx}

\usepackage{amsmath}
\usepackage{graphicx}
\usepackage{wrapfig}
\usepackage{sidecap}
\usepackage{placeins}
\usepackage{tikz}
\usepackage{bbm}
\usetikzlibrary{cd}

\def\bsymb{\boldsymbol}
\def\cM{{\cal M}}
\def\cX{{\cal X}}
\def\cJ{{\cal J}}
\def\cN{{\cal N}}
\def\cP{{\cal P}}
\def\bx{{\mathbf x}}
\def\bR{{\mathbf R}}

\def\bpi{{\bsymb \pi}}
\def\E{\mathbb{E}}
\def\H{\mathbb{H}}
\def\R{\mathbb{R}}
\newcommand{\nn}{\nonumber}

\def\bpprior{\bpi_{\text{prior}}}

\def\inv{^{-1}}

\def\inv{^{-1}}

\def\bcX{{\boldsymbol \cX}}
\newcommand{\PP}{\mathrm{P}}
\newcommand{\RR}{\mathrm{R}}

\DeclareMathOperator*{\argmin}{argmin}

\title{Characterizing optimal hierarchical policy inference on graphs
via non-equilibrium thermodynamics}

\author{
  Daniel McNamee \\
  Computational and Biological Learning Lab, University of Cambridge\\
  \texttt{d.mcnamee@eng.cam.ac.uk}
}

\begin{document}

\maketitle

\begin{abstract}
Hierarchies are of fundamental interest in both stochastic optimal control and biological control due to their facilitation of a range of desirable computational traits in a control algorithm and the possibility that they may form a core principle of sensorimotor and cognitive control systems. However, a theoretically justified construction of state-space hierarchies over all spatial resolutions and their evolution through a policy inference process remains elusive. Here, a formalism for deriving such normative representations of discrete Markov decision processes is introduced in the context of graphs. The resulting hierarchies correspond to a hierarchical policy inference algorithm approximating a discrete gradient flow between state-space trajectory densities generated by the prior and optimal policies.
\end{abstract}

\section{Stochastic policy inference in discrete Markov decision processes}

Let $\cM:=\{\cX,\PP,\RR\}$ be a discrete MDP where states are denoted $x\in\cX$ and $\RR(x)$ is the reward associated with a state $x$. It is assumed that actions are deterministic transitions on the state-space $\cX$ and thus the MDP can be summarized as a directed graph (though the general formalism is not restricted to this case). Let $\bx:=(x_0,\ldots,x_T)$ be a state trajectory\footnote{In general, bold type will be used when referring to trajectories.} with an arbitrary horizon $T$ (which might be infinite). Then the MDP objective from a ``sum-over-paths'' perspective \cite{Kappen2005} is
\begin{equation}
    \bpi^* := \argmin_\bpi \E_{\bpi} \left[-\bR(\bx)\right] \equiv \argmin_\bpi \E_{\bpi} \left[-\sum_{t=0}^T \RR(x_t)\right] \label{eqn:MDPobj}
\end{equation}
where $\bpi$ is the ``policy'' over state trajectories induced from the policy $\pi$. This MDP can be embedded in a linearly solvable MDP \cite{Todorov2007} (or equivalently, a KL Control problem \cite{Kappen2012}) by subtracting a policy description length penalty $-\beta\inv \log \bpi(\bx)$ for each state trajectory $\bx$ \cite{Rawlik2012}. This allows the objective (Eqn.~\ref{eqn:MDPobj}) to be re-formulated as a free energy functional $\cJ[\bpi]$ incorporating a policy entropy $\H[\bpi]$ maximization objective:
\begin{eqnarray}
\cJ[\bpi] &=& \E_{\bpi} \left[-\bR(\bx) + \beta\inv\log{\bpi(\bx)}\right] \nn \\
&=& E\left[\bpi\right] + \beta\inv S\left[\bpi\right] \label{eqn:free_energy}
\end{eqnarray}
where the entropy and energy terms are denoted $E:=\E_{\bpi} \left[-\bR(\bx)\right]$ and $S:=-\H[\bpi]$ respectively. This policy functional encapsulates the trade-off between policy value and policy complexity (Eqn.~\ref{eqn:free_energy}) in the space of state trajectories $\bcX$. By considering this inverse temperature parameter in the limit $\beta\rightarrow \infty$, one recovers the standard MDP objective of maximizing expected cumulative reward \cite{Rawlik2012,Theodorou2013}. For simplicity, the inverse temperature is set to $\beta=1$. A well-known trick \cite{Mitter2003,Todorov2007}, is that both the optimal trajectory policy $\bpi^*$ and the minimum energy $\cJ[\bpi^*]$ can be expressed in closed form using the Gibbs-Boltzmann distribution:
\begin{eqnarray}
\bpi^*(\bx) &\propto& \bpprior e^{\bR(\bx)} \nn \\
\cJ[\bpi^*] &=& -\log \E_{\bpprior} \left[ e^{\bR(\bx)}\right] \label{eqn:equilibrium}
\end{eqnarray}
These formulae explicitly reflect the balance between a control complexity penalty (minimized by the prior policy $\bpprior$) and reward $e^{\bR(\bx)}$. An optimal transition policy $\pi^*$ can be computed by marginalizing over future trajectories, $\pi^*(x_{t+1}|x_t)=\sum_{\bx_{t+2:T}}\bpi^*(\bx_{t+1:T}|x_t)$, a calculation to which standard approximate inference methods can be applied \cite{Kappen2012}. However, the goal of this study is not to compute a transition policy $\pi$ but to gain insight into the hierarchical structure of the MDP. This will be accomplished by characterizing the dynamics of the trajectory policy $\bpi$ generated by an optimal policy inference algorithm as it is transformed from the prior $\bpprior$ to the optimal policy $\bpi^*$.

\section{Planning via non-equilibrium thermodynamics}

This question will be addressed by exploiting the physical interpretation of the free energy functional (Eqn.~\ref{eqn:free_energy}). It is the energy functional of a stochastic system of ``particles'' taking ``positions'' in the discrete space of state trajectories which (i) are initialized accord to the density $\bpprior$, and (ii) evolve over time $\tau$ under the influence of the potential $\Psi(\bx):=-R(\bx)$.  The Langevin dynamics associated with the potential $\Psi(\bx)$ cause particles to be attracted to state trajectories $\bx$ with higher rewards. The trajectory policy $\bpi_\tau$ reflects the density of such particles in trajectory space at time $\tau$. The unique thermal equilibrium of this system (i.e. the density of particles in the lowest energy state) is the optimal policy $\bpi^*\equiv \bpi_\infty$ given by Eqn.~\ref{eqn:equilibrium}. Conceptually, planning (as inference \cite{Toussaint2009}) is accomplished by initializing the particle system at the prior policy $\bpprior$ (away from the equilibrium density $\bpi^*$) and then allowing the particles to evolve according to the non-equilibrium thermodynamics described above. Instead of simulating the stochastic dynamics of these particles, the evolution of their density is studied. This is given by the Fokker-Planck (FP) equation for which the optimal policy $\bpi^*$ is the unique stationary solution \cite{Ambrosio2008}:
\begin{equation}
    \dot{\bpi} = \nabla \left[(\nabla \Psi)\bpi \right] + \Delta \bpi \label{eqn:FP}
\end{equation}
where $\bpi:[0,\infty)\times \bcX \rightarrow \R$ doubles notationally as a \emph{policy flow} (i.e. at each policy inference timepoint $\tau$, $\bpi(\tau,\cdot):=\bpi_\tau$ defines a trajectory density). Furthermore, the FP equation has the following remarkable property -- the time evolution of $\bpi_\tau$ according to the FP equation is guaranteed to reach the optimal policy $\bpi_\tau\rightarrow \bpi^* ,\tau\rightarrow \infty$ by the steepest descent\footnote{With respect to the Wasserstein metric -- see the supplementary information for a definition.} on the free energy functional $\cJ$ \cite{Jordan1998,Chow2012,Maas2011}. At this global minimum, $\bpi_\infty$ will equal the optimal policy $\bpi^*$ and the particles will remain in equilibrium.  This implies that the FP equation is the \emph{gradient flow} of the free energy $\cJ[\bpi]$, and therefore the evolution of $\bpprior\rightarrow \bpi^*$ described by equation~\ref{eqn:FP} is that generated by an optimal policy inference algorithm. Since this stochastic process is considered at the level of trajectories, it is described as ``hierarchical'' policy inference with respect to state transitions since it is sensitive to dependencies between states at all temporal horizons. In the next section, the optimal probability density flux between trajectories (Eqn.~\ref{eqn:discrete_FP_plan}) will be re-expressed as a function of states (Eqn.~\ref{eqn:flux_states}). Summing the total flux at each state (Eqn.~\ref{eqn:main_result}) at each planning timepoint $\tau$ returns a dynamic state hierarchy reflecting the magnitude of policy flux at each state under optimal policy flow.

\section{Policy flow as a discretized differential equation}

Applying the FP equation (Eqn.~\ref{eqn:FP}) in the context of variational optimization requires a temporal discretization \cite{Jordan1998,Ambrosio2008}. Here, a spatial discretization is also required since the MDP state-spaces considered are discrete. A spatially discretized form of the Fokker-Planck equation can be constructed based on a Markov chain approximation of the underlying Langevin dynamics such that it retains the key properties of the continuous case \cite{Chow2012,Maas2011}. Let $\bpi$ be the current trajectory policy (e.g. $\bpi=\bpprior$ at planning timestep $\tau=0$) and define the \emph{generalized potential} $\bar{\Psi}(\bx):=-\bR(\bx)+\log{\bpi(\bx)}$. Then the discrete FP equation (Eqn.~\ref{eqn:discrete_FP}, Supplementary Information) is
\begin{eqnarray}
\dot{\bpi}(\bx)
&=& \sum_{\bx'\in\bcX} K(\bx,\bx')M(\bx,\bx')\left[\bar{\Psi}(\bx')-\bar{\Psi}(\bx)\right] \nn \\
K(\bx,\bx') &=& \begin{cases}
      \bpi(\bx') & \text{if}\ \bar{\Psi}(\bx')>\bar{\Psi}(\bx) \\
      \bpi(\bx) & \text{if}\ \bar{\Psi}(\bx')<\bar{\Psi}(\bx) \\
    \end{cases} \label{eqn:discrete_FP_plan}
\end{eqnarray}
and $M$ is the Markov stochastic matrix over the set of trajectories generated by the current policy and $K$ is a Laplacian kernel. Since trajectories are conditionally independent (across ``samples'' or ``episodes'' in the MDP),  $M(\bx,\bx')=\pi(\bx')$. If $\bar{\Psi}(\bx')>\bar{\Psi}(\bx)$, the policy density at $\bx$ increases due to the inflow of policy ``particles'' attracted by the lower energy state associated with the generalized potential $\bar{\Psi}(\bx)$. The policy flux along the ``velocity field'' defined by $K$ is driven by two fluxes (i) $-\nabla \bR$ ``pulls'' policies towards action sequences with large expected rewards, and (ii) $\nabla \log{\bpi}$ ``pushes'' the policy away from policies with large description lengths. Summing over trajectories, the total flux is
\begin{eqnarray}
\sum_{\bx\in\bcX} \dot{\bpi}(\bx) &=& \sum_{\bx,\bx'\in\bcX} K(\bx,\bx')M(\bx,\bx') \left[\bar{\Psi}(\bx')-\bar{\Psi}(\bx) \right] \\
&=& \sum_{\bx,\bx'\in\bcX} \bpi(\bx)\bpi(\bx') \left[\Gamma^-(\bx,\bx') - \Gamma^+(\bx',\bx)\right]
\end{eqnarray}
where the following functions have been used:
\begin{eqnarray}
\Gamma^+(\bx,\bx') &:=&
\begin{cases}\bar{\Psi}(\bx')-\bar{\Psi}(\bx) & \text{if } \bar{\Psi}(\bx')>\bar{\Psi}(\bx) \\ 0 & \text{ otherwise} \end{cases} \\
  \Gamma^-(\bx,\bx') &:=&
  \begin{cases}\bar{\Psi}(\bx)-\bar{\Psi}(\bx') & \text{if } \bar{\Psi}(\bx')<\bar{\Psi}(\bx) \\ 0 & \text{ otherwise} \end{cases} ~~.
\end{eqnarray}
These functions measure the density gain $\Gamma^+(\bx,\bx')$ and loss $\Gamma^-(\bx,\bx')$ at each trajectory $\bx$ due to the optimal policy flux with respect to another trajectory $\bx'$. Note that, for every $\Gamma^+(\bx,\bx')$ term there is an equal and opposite $\Gamma^-(\bx',\bx)$ term and therefore $\sum_{\bx} \dot{\bpi}(\bx)=0$ reflecting the FP equation's role as a differential continuity equation for the conservation of total probability i.e. $\sum_\bx \dot{\bpi}(\bx)=0$.

\section{``Pulling back'' from trajectories to transitions}

Although the total flux across trajectories is zero, the total flux associated with a particular \emph{state} in the original state-space $\cX$ is not and reveals a state-space hierarchy. In this section, formulae are recorded for these total flux values which reflect a state's ``rank'' in the optimal hierarchy (Section~\ref{sec:transexp}, Supplementary Information). For a pair $(\bx,\bx')$ of trajectories such that $\bar{\Psi}(\bx')>\bar{\Psi}(\bx)$:
\begin{eqnarray}
    \Gamma^+(\bx,\bx') &=& \sum_{t=0}^{|\bx|-1}\sum_{t'=0}^{|\bx'|-1} \left[\Gamma(x_t,x'_{t'}) \right] \\
    \Gamma(x_t,x'_{t'}) &=& \bar{\Psi}(x'_{t'+1}|x'_{t'})-\bar{\Psi}(x_{t+1}|x_t)
  \end{eqnarray}
where $\bar{\Psi}(x_{t+1}|x_t) := \log{\left[\pi(x_{t+1}|x_t)e^{-R(x_{t})} \right]}$ is the contribution to the generalized potential from transition $x_t\rightarrow x_{t+1}$. Using Eqn.~\ref{eqn:sumofpairs}, one can rewrite this in terms of states as:
\begin{eqnarray}
  \sum_{\bx\in\bcX} \dot{\bpi}(\bx) &=& \sum_{x,x'\in\cX}K_D(x,x') \left[\cJ_\pi(x) - \cJ_\pi(x') \right] \label{eqn:flux_states}
\end{eqnarray}
where $\cJ_\pi(x) = \E_{x'\sim\pi(x'|x)}[\bar{\Psi}(x'|x)]=-R(x) - \H_\pi(x)$ is the local contribution to the free energy objective (Eqn.~\ref{eqn:free_energy}) and
\begin{eqnarray}
K_D(x,x') &=& \begin{cases}
      D_{x_0,x'}D_{x'x} & \text{if}\ \cJ_\pi(x') > \cJ_\pi(x) \\
      D_{x_0,x}D_{xx'} & \text{if}\ \cJ_\pi(x') < \cJ_\pi(x)
    \end{cases} ~~,
\end{eqnarray}
where $x_0$ is the initial state and $D:=(I-P_\pi)^{-1}$ is the fundamental matrix of the Markov chain generated by the policy $\pi$. Ultimately, the discrete Fokker-Planck equation can be re-expressed geometrically using a discrete Laplacian operator $\Delta_D$ defined by $K_D$ \cite{Wardetzky2007}:
\begin{equation}
  \dot{\bpi}(x) = \Delta_D \cJ_\pi(x) ~~.  \label{eqn:main_result}
\end{equation}
The role of the discrete Laplacian operator here has an intuitive physical meaning. The generalized potential is not independent across states $x$ (due to its dependence on $\pi$) analogous to the dependence of, for example, a gravitational potential across distinct positions in the real world. A typical operation in mathematical physics is to identify the independent sources which generate a physical field (e.g. planets). This is accomplished by taking the divergence of the field which is equivalent to the Laplacian of the field's potential. Considering the local energy potential $\cJ_\pi(x)$ as generating a ``planning complexity field'', this suggests that equation \ref{eqn:main_result} can be interpreted as FP dynamics ``solving'' the planning problem by targeting the independent sources which ``generate'' the planning complexity.

\section{Example: a regular graph}

\begin{figure}[hb!]
\centering
\includegraphics[width=\textwidth]{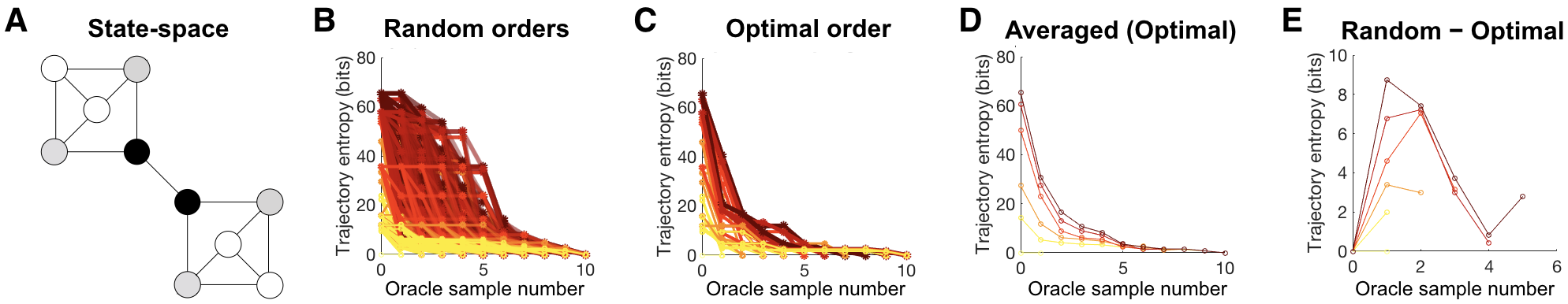}
\caption{Trajectory entropy dynamics due to ``oracle'' planning on a regular graph.}
\label{fig:example}
\end{figure}

As a simple, expository application, a regular graph which has previously been used for behavioral experimentation \cite{Solway2014} is considered. In Fig.~\ref{fig:example}A, each node represents a state (displayed as a fractal to the participants in the experiment) and each edge an available transition. Regular graphs such as this provide a challenging scenario for elucidating hierarchy due to their homogeneous local structure (the number of available transitions is the same at each state). Despite this, there is an obvious bottleneck between two ``rooms'' in Fig.~\ref{fig:example}A. Participants were tasked with solving shortest-path problems between uniformly drawn states. In Fig.~\ref{fig:example}A, darker node color reflects higher $\Delta_D \cJ_\pi(x)$ (Eqn.~\ref{eqn:main_result}) and thus one can observe that the key bottleneck is recovered. Participants were also asked to identify ``bottleneck'' states in this state-space despite never having observed its global structure. The hierarchical state ranking measure $\Delta_D \cJ_\pi(x)$ approximately matched the full distribution of bottlenecks chosen explicitly by subjects \cite{Solway2014} which had not been previously predicted in a model.

The implication of the analysis presented here is that an optimal planning process should prioritize the identification of transitions in the hierarchical order measured by $\Delta_D \cJ_\pi(x)$ as this reflects the steepest descent on $\cJ[\bpi]$ over all trajectories. From an information-theoretic perspective, a transition at a bottleneck state ``communicates'' the most amount of information regarding a shortest path between states on average and the second highest ranked states give the next highest amount of information and so on. This prediction was tested by simulating an ``oracle planner'' which simply identifies the optimal transition at a given state on a shortest path or reveals that the state is not on the shortest path. This procedure iteratively refines the set of candidate trajectories until a shortest path is identified. The entropy of the remaining candidate trajectories measures the expected amount of information remaining to be specified by the planning algorithm. This is plotted as a function of the number of oracle samples in Fig.~\ref{fig:example}B-E. In Fig.~\ref{fig:example}C, trajectory entropy is plotted for the optimal oracle sampling order, and in Fig.~\ref{fig:example}D, the average per shortest path length (darker is longer) is presented. As expected, no random oracle sampling order (Fig.~\ref{fig:example}B) outperforms the optimal hierarchical sampling order on average across tasks (Fig.~\ref{fig:example}E).

\vspace{-0.5em}
\vspace{-1em}

\section{Discussion}

A preliminary study of discrete gradient flow in the context of hierarchical policy inference has been presented. Amongst all possible policy inference processes, the FP equation describes the expected time evolution of those which take the steepest descent on the free energy objective thus establishing a principle of optimality. Considering policy inference in trajectory space implies that state-space structure at all scales is integrated into the policy inference process. The resulting policy flow provides a normative perspective on optimal hierarchical policy inference however it remains to be seen whether the principles established here can be usefully embedded within practical algorithms. The basic example provided establishes a link with hierarchical representations in human cognition and thus a potentially fruitful research direction will be to examine whether aspects of the behavioral and neural dynamics of natural planning and problem-solving can be explained within this framework.

\newpage

\subsection*{Acknowledgements}

Thanks to the reviewers for their feedback and M\'{a}t\'{e} Lengyel and Daniel Wolpert for discussions.
\bibliographystyle{plainnatnourl}
\bibliography{library}


\newpage
\normalsize

\part*{Supplementary Information}






\section{Free energy, the Fokker-Planck equation, and variational optimization}

A brief, self-contained, introduction to the properties of the Fokker-Planck equation in the context of variational optimization is provided.

\subsection{Stochastic processes and the Fokker-Planck equation}

A general class of stochastic processes is given by the It\^{o} stochastic differential equation:
\begin{equation}
dx(t) = \mu(x(t),t)\,dt + \sqrt{2}\sigma(x(t),t)\, dw(t) \label{eqn:ito}
\end{equation}
This equation describes the evolution of a random vector $x\in X$ under a \emph{drift} $\mu(x(t),t)\,dt$ and a \emph{diffusion} $\sigma(x(t),t)\, dw(t)$. The diffusion coefficient $\sigma(x(t),t)$ scales a Wiener process $w(t)$ which describes the accumulation of gaussian noise $w(t+dt)-w(t)\sim \cN(0,dt)$. In general $x_t$ and $\mu(x(t),t)$ are $n$-dimensional random vectors, $\sigma(x(t),t)$ an $(n,m)$-dimensional matrix, and $w(t)$ an $m$-dimensional vector. The \emph{Fokker-Planck equation} describes the time-evolution of the probability density $\rho$ of the random vector $x$:
\begin{equation}
\dot{\rho} = -\nabla (\mu \rho) +\Delta (\sigma^2 \rho) \label{eqn:fp}
\end{equation}
where $\rho:[0,\infty)\times M \rightarrow [0,\infty)$ is a probability density flow (i.e. at each timepoint $t$, $\rho(t,\cdot)$ defines a density), $M$ is the underlying manifold, and $\sigma^2:=\sigma \sigma^T$ is known as the \emph{diffusion tensor}. The simplest case of the FP equation is Brownian motion with stationary (i.e. time-independent) and invariant (i.e. position-independent) diffusion $\sigma(x(t),t)=\sqrt{\beta^{-1}}$ and zero drift $\mu(x(t),t)=0$. In this case, the FP equation (\ref{eqn:fp}) is the \emph{heat equation} with inverse temperature $\beta$ controlling the ``speed'' of the diffusion:
\begin{equation}
\dot{\rho}= \beta^{-1} \Delta \rho \label{eqn:heat}
\end{equation}

\subsection{Langevin dynamics: drift potential}

The special case in which the drift $\mu$ is a conservative time-homogeneous vector field corresponds to \emph{Langevin dynamics}. Here the vector field $\mu$ is generated by a scalar potential $\mu(x(t),t)=-\nabla \Psi(x(t))$. Further assuming a ``white noise'' diffusion with temperature $\beta$ results in the stochastic dynamics:
\begin{equation}
dx(t) = \nabla \Psi(x(t))\,dt + \sqrt{2\beta^{-1}}\, dw(t) \label{eqn:itoJKO}
\end{equation}
In this case, the FP equation (Eqn.~\ref{eqn:fp}) becomes
\begin{equation}
\dot{\rho} = \nabla\cdot \left[(\nabla \Psi ) \rho\right] + \beta^{-1}\Delta \rho \label{eqn:fpJKO}
\end{equation}
and the corresponding velocity field $v$ is
\begin{equation}
v = -\nabla \Psi - \beta^{-1}\nabla\left(\log \rho\right)
\end{equation}
such that $\dot{\rho} = -\nabla (\rho v)$. The expression $j:=\rho v$ is referred to as the \emph{current density} which reflects the rate of probability density flow at a particular point $x\in X$.

\subsection{Free energy minimization}

The FP equation is intimately related to an associated free energy functional $\cJ$ of the density $\rho$:
\begin{eqnarray}
\cJ[\rho] &=& \E_{\rho} \left[\Psi + \beta\inv\log{\rho}\right] \nn \\
&=& E\left[\rho\right] + \beta\inv S\left[\rho\right] \label{eqn:free_energy_supp}
\end{eqnarray}
where we have the \emph{energy} $E:=\E_{\rho} \left[\Psi\right]$ and \emph{entropy} $S:=-\H[\rho]$ terms respectively. The free energy $\cJ$ is a Lyapunov function for the FP dynamics (Eqn.~\ref{eqn:fpJKO}). Conversely, and remarkably, the FP equation is a \emph{gradient flow} of the free energy functional \cite{Jordan1998}. Specifically, the FP equation defines the trajectory of steepest descent on the free energy $\cJ$ with respect to the 2-Wasserstein metric $d_{W_2}$ on the space $\cP(X)$ of densities $\rho,\rho'\in \cP(X)$ on $X$:
\begin{equation}
d_{W_2}(\rho,\rho') := \inf_{\lambda\in M(\rho,\rho')} \int_{X\times X} d(x,x')^2 d\lambda
\end{equation}
where $M(\rho,\rho')$ is the set of probability densities $\lambda$ on $X\times X$ whose marginals are equal to $\rho$ and $\rho'$ respectively. Although no derivations in this work will rely on the definition of the Wasserstein metric, it may be interesting to point out that the appearance of this metric at this juncture is related to its theoretical foundations in optimal transport theory \cite{Villani2009}.

\subsection*{Discrete Fokker-Planck theory}

Here, our primary interest lies in discrete Markov decision problems thus the translation of continuous FP theory to the discrete domain is reviewed \cite{Chow2012,Maas2011}. $X$ is a discrete state-space (e.g. a graph) with states $x\in X$. A density on this state-space is a discrete distribution $\rho:X\rightarrow [0,1]$ such that $\sum_{x\in X}\rho(x) = 1$. A discrete potential $\Psi$ is defined as a mapping from states to real numbers $\Psi:X\rightarrow \R$. The discrete form of the Fokker-Planck equation (Eqn.~\ref{eqn:discrete_FP}) is based on a Markov chain approximation of Langevin stochastic dynamics and retains the key properties of the continuous case \cite{Chow2012,Maas2011}:
\begin{eqnarray}
\dot{\rho}(x)
&=& \sum_{x'\in\cN(x)} K(x,x')\left[\Psi(x')-\Psi(x)+\log{\rho(x')}-\log{\rho(x)}\right] \nn \\
&=& \sum_{x'\in\cN(x)} K(x,x')\left[\bar{\Psi}[\rho](x')-\bar{\Psi}[\rho](x)\right] \nn \\
&=& \Delta \bar{\Psi}[\rho](x) \nn \\
K(x,x') &=& \begin{cases}
      \rho(x') & \text{if}\ \bar{\Psi}(x')>\bar{\Psi}(x) \\
      \rho(x) & \text{if}\ \bar{\Psi}(x')<\bar{\Psi}(x)
    \end{cases} \label{eqn:discrete_FP}
\end{eqnarray}
where $\bar{\Psi}$ is the \emph{generalized potential}:
\begin{equation}
\bar{\Psi}(x) := \Psi(x) + \beta^{-1} \log{\rho(x)} \label{eqn:genpotential}
\end{equation}
The function $K$ can be interpreted as the kernel of a discrete Laplacian operator $\Delta$ applied to the generalized potential $\bar{\Psi}$ \cite{Wardetzky2007}.

\section{From trajectories to transitions: expectations of linear functions}
\label{sec:transexp}
Results are derived which will be used in the derivations in the main paper. An absorbing\footnote{Analogous results can be derived for regular Markov chains.} Markov chain with stochastic matrix $\pi$ is considered. Let $f(\bx)$ be a function of state trajectories $\bx\in\bcX$ in this chain and $\bpi$ be the induced probability distribution over trajectories. Assume that $f$ can be expressed as a sum over states:
\begin{equation}
  f(\bx) := \sum_{i=1}^{T} f(x_i) ~~. \nn \\
\end{equation}
It is known that the expected value of $f$ is \cite{Dayan1993}
\begin{eqnarray}
\E_\bpi[f]   &=& \sum_{x\in\cX} D_{x_0 x}f(x) ~~,
\end{eqnarray}
where $D$ is the fundamental matrix of the Markov chain defined by $\pi$ \cite{Kemeny1983} and $x_0$ is the initial state.

Analogous results for transitions are required. Assume that $f$ can be expressed as a sum over transitions
\begin{equation}
  f(\bx) := \sum_{t=0}^{T-1} f(x_i,x_{i+1})
\end{equation}
Then
\begin{eqnarray}
\E_\bpi[f] &:=& \sum_{\bx\in\bcX} \bpi(\bx)f(\bx) \nn \\
&=& \sum_{T>0}\sum_{\bx^T\in\bcX^T} \bpi(\bx^T) \sum_{t=0}^{T-1}f\left(x^T_t,x^T_{t+1}\right) \nn \\
&=&  \sum_{T>0} \sum_{x,x'\in\cX}\sum_{t=0}^{T-1} p^t_\pi(x|x_0) \pi(x'|x) f(x,x')
\end{eqnarray}
where $p^t_\pi(x|x_0)$ is the probability of being in state $x$ on timestep $t$ after starting in state $x_0$.
Therefore
\begin{eqnarray}
        \E_\bpi[f] & = & \sum_{x,x'\in\cX} D_{x_0 x}\pi(x'|x)f(x,x') \nn \\
        &=& \sum_{x\in\cX} D_{x_0 x}\E_{x'\sim \pi(x'|x)}\left[f(x,x') \right] \label{eqn:exp_trans}
\end{eqnarray}

An intuitive example is given by the transition identity function $f:=\mathbbm{1}_{x\rightarrow x'}$:
\begin{equation}
\mathbbm{1}_{x\rightarrow x'}(y,y')=
\begin{cases}
1 & \text{iff } x=y,\ x'=y' \\
0 & \text{otherwise}
\end{cases}
\end{equation}
Then $\mathbbm{1}_{x\rightarrow x'}(\bx)$ simply counts the number of occurences of transition $x\rightarrow x'$ in trajectory $\bx$ and
\begin{eqnarray}
\E_\bpi[\mathbbm{1}_{x\rightarrow x'}] &=& D_{x_0 x} \pi(x'|x)
\end{eqnarray}
which, as expected, is simply the expected number of times the chain is in state $x$ after starting in state $x_0$ multiplied by the probability of transitioning to $x'$. Similarly, letting $f$ be the state identity function $f(x):=\mathbbm{1}_x$ implies that $E_\bpi[f]$ returns an entry of the fundamental matrix which equals the expected number of times state $x$ is encountered:
\begin{eqnarray}
\E_\bpi[\mathbbm{1}_x]   &=& D_{x_0 x} ~~.
\end{eqnarray}

Applying discrete Fokker-Planck dynamics requires us to develop analogous arguments for pairs of trajectories. Let $f(\bx,\bx')$ be a linear sum of states $x,x'\in\cX$ in trajectories $\bx,\bx'\in\bcX$. The contribution $f(x,x')$ to $f(\bx,\bx')$ for a specific state pair $(x,x')$ is weighted by the multiplicity of the state combination $(x,x')$ in the set of trajectories $\bcX$ according to the distribution $\pi(\bx,\bx')=\bpi(\bx)\bpi(\bx')$. Analogous to previous arguments, this is equivalent to taking the expectation with respect to the probability of generating a trajectory according to $\pi$ with $x$ and then $x'$ appearing on arbitrary timesteps:
\begin{eqnarray}
\E_\bpi[f] &=& \sum_{\bx,\bx'\in\bcX} \bpi(\bx)\bpi(\bx')f(\bx,\bx') \nn \\
&=&  \sum_{x,x'\in\cX}\sum_{T,T'>0}\sum_{t=0}^{T-1} \sum_{t'=0}^{T'-1} p^t_\pi(x|x_0)p^{t'}_\pi(x'|x) f(x,x') \nn \\
&=& \sum_{x,x'\in\cX} D_{x_0 x}D_{xx'}f(x,x')
\end{eqnarray}
Finally, we consider the case where $f(\bx,\bx')$ decomposes as a sum of transition pairs within each trajectory:
\begin{eqnarray}
  f(\bx,\bx') = \sum_{x,y,x',y'\in\cX} f(x,y) + f(x',y')
\end{eqnarray}
Then
\begin{eqnarray}
  \E_\bpi[f] = \sum_{x,x'\in\cX}D_{x_0 x}D_{xx'} \left\{\E_{\pi(y|x)}\left[f(x,y)\right] + \E_{\pi(y'|x')}\left[f(x',y') \right] \right\} \label{eqn:sumofpairs}
\end{eqnarray}

\end{document}